\documentclass{article}
\usepackage{amsfonts}

\def\noi{\noindent}

\makeatletter

\renewcommand{\thesubsubsection}%
        {\arabic{section}.\arabic{subsection}.\arabic{subsubsection}.}
\renewcommand{\@oddhead}{\raisebox{0pt}[\headheight][0pt]{%
   \vbox{\hbox to\textwidth{\rightmark \hfil \rm \thepage \strut}\hrule}}}
\renewcommand{\@evenhead}{\raisebox{0pt}[\headheight][0pt]{%
   \vbox{\hbox to\textwidth{\thepage \hfil \leftmark \strut}\hrule}}}
\newcommand{\heads}[2]{\markboth{\protect\small\it #1}{\protect\small\it
#2}}
\newcommand{\Acknow}[1]{\subsection*{Acknowledgement} #1}
\makeatother

\newcommand{\Title}[1]{\noi {\Large #1} \\}
\newcommand{\Author}[2]{\noi{\large\bf #1}\\[2ex]\noi{\it #2}\\}
\newcommand{\Abstract}[1]{\vskip 2mm \begin{center}
     \parbox{16.4cm}{\small\noi #1} \end{center}\bigskip}
\newcommand{\foom}[1]{\protect\footnotemark[#1]}
\newcommand{\foox}[2]{\footnotetext[#1]{#2}
		\addtocounter{footnote}{1}}
\newcommand{\email}[2]{\footnotetext[#1]{e-mail: #2}
		\addtocounter{footnote}{1}}
\def\sect{Sec.\,}
\def\Ref{Ref.\,\cite}

\def\nqq{\hspace{-2em}}
\def\nhq{\hspace{-0.5em}}

\def\cm{\hspace{1cm}}

\def\eq{Eq.\,}
\def\eqs{Eqs.\,}
\def\beq{\begin{equation}}
\def\eeq{\end{equation}}
\def\bear{\begin{eqnarray}}
\def\al{&\nhq}
\def\lal{&&\nqq {}}               
\def\bearr{\bear \lal}
\def\ear{\end{eqnarray}}

\def\tst{\textstyle}
\def\dst{\displaystyle}

\def\nn{\nonumber\\ {}}

\def\eql{\al =\al}

\def\eqdef{\stackrel{\rm def}{=}}
\def\e{{\,\rm e}}
\def\d{\partial}

\def\sign{\mathop{\rm sign}\nolimits}
\def\diag{\mathop{\rm diag}\nolimits}
\def\dim{\mathop{\rm dim}\nolimits}
\def\const{{\rm const}}

\def\half{{\tst\frac{1}{2}}}

\newcommand{\vars}[1]{\left\{\begin{array}{ll}#1\end{array}\right.}

\def\wide {\vphantom{\dst\int}}

\def\cR{{\cal R}}
\def\ocR{\overline{\cR}}
\def\R{{\mathbb R}}
\def\M{{\mathbb M}}
\def\V{{\mathbb V}}
\def\mN{_M^N}
\def\og{{\overline g}}
\def\uc{{\underline c}}
\def\vY{{\vec Y}}
\def\oo{\omega_1}

\def\eom{\eta_\omega}

\def\dsJ{\mbox{$ds^2_{\rm J}$}}
\def\fin{\mathop{\rm fin}\nolimits}
\def\cf{c_{\varphi}}
\def\etaF{\eta_{{}_F}}

\heads{K.A. Bronnikov and J.C. Fabris}
      {Nonsingular multidimensional cosmologies without fine tuning}

\begin{document}
\thispagestyle{empty}
\begin{flushright}
                                         	{\bf hep-th/0207213}
\end{flushright}
\medskip

\Title{\bf Nonsingular multidimensional cosmologies without fine tuning}

\Author{K.A. Bronnikov\foom 1 and J.C. Fabris\foom 2}
{Departamento de F\'{\i}sica, Universidade Federal do Esp\'{\i}rito Santo,
Vit\'oria, CEP29060-900, Esp\'{\i}rito Santo, Brazil}

\Abstract
{Exact cosmological solutions for effective actions in $D$ dimensions
inspired by the tree-level superstring action are studied. For a certain
range of free parameters existing in the model, nonsingular bouncing
solutions are found. Among them, of particular interest can be open
hyperbolic models, in which, without any fine tuning, the internal scale
factor and the dilaton field (connected with string coupling in string
theories) tend to constant values at late times. A cosmological singularity
is avoided due to nonminimal dilaton-gravity coupling and, for $D > 11$, due
to pure imaginary nature of the dilaton, which conforms to currently
discussed unification models. The existence of such and similar solutions
supports the opinion that the Universe had never undergone a stage driven by
full-scale quantum gravity.  }


\foox 1 {Permanent address: VNIIMS, 3-1 M. Ulyanovoy St., Moscow 117313,
	Russia, {\it and\/} Institute of Gravitation and Cosmology,
	RUDN, 6 Miklukho-Maklaya St., Moscow 117198, Russia; e-mail:
        kb@rgs.mccme.ru}

\email 2 {fabris@cce.ufes.br}

\section{Introduction}   

    It is widely recognized that the early Universe at sub-Planckian scales
    has been a scene for many strong effects which are nowadays either
    weak or even unobservable. Even remaining on a semiclassical level, one
    has to take into account the quantum properties of matter, the probable
    dynamical nature of extra dimensions predicted by modern unification
    theories (strings, supergravities, etc.), the existence of unusual kinds
    of matter such as strings and/or branes as well as modification of
    gravity at high energy densities and space-time curvatures.


    Many of these features show the existence of semiclassical and even
    classical (tree-level) mechanisms able to circumvent the well-known
    singularity theorems and to prevent the formation of a cosmological
    singularity, keeping the curvature on sub-Planckian scales. This means
    that very probably there has not been an epoch of cosmological evolution
    driven by full-scale quantum gravity.

    Thus, some nonsingular models have been built with the aid of
    gravitational Lagrangians nonlinear in curvature (e.g., \cite{star,
    brand99}), whose origin may be explained by the quantum properties of
    matter fields. Supergravities and string theories, the candidate
    ``theories of everything", also suggest new opportunities.
    Though, if the extra space-time dimensions, being an inevitable
    ingredient in such theories, are considered dynamically, the singularity
    problem becomes even more involved since, in addition to the usual
    cosmological scale factor, the extra dimensions can collapse or blow up,
    leading to a curvature singularity.

    In this paper we will deal with particular multidimensional
    cosmologies obtainable from an effective action which conforms to the
    low energy limit of some currently discussed unification theories ---
    see \cite{duff,hukhu} and references therein. (For recent reviews on
    string cosmologies see \cite{lidsey, gasvenez}.)
    This effective action admits as many as three natural
    mechanisms able to violate the usual energy conditions and hence
    potentially lead to nonsingular cosmological models. One mechanism
    is related to the nonminimal nature of the dilatonic scalar field
    and its interaction with antisymmetric form fields. Similar models
    in 4 dimensions have been discussed in \Ref{fabris}. Second, the
    Brans-Dicke coupling constant $\omega$ may have values leading to a
    ``wrong'' sign of the dilaton kinetic term in the Einstein-frame
    Lagrangian, which happens in dimensions larger than 11 \cite{khve} (in
    other terms, the dilaton becomes pure imaginary). Third, in some
    field models of string origin, formulated in space-times with multiple
    time coordinates, antisymmetric forms can have Lagrangians with a
    ``wrong" sign \cite{hukhu}.

    The models to be discussed are quite simple but natural from the
    viewpoint of the underlying theories. In a sense, our ``vacuum" approach
    is alternative to that of \Ref{brand00} where a hot brane gas is
    considered whereas global antisymmetric forms are ignored.  We shall see
    that the first two mechanisms (but not the third one) really work and
    lead to globally regular cosmological solutions to the field equations.

    Our aim here is only to demonstrate the effect of these mechanisms,
    therefore we do not consider the totality of exact solutions
    that might be obtained for the action (\ref{SJ})
    but restrict our attention to the simplest case: a single dilatonic field
    with a Brans-Dicke type Lagrangian, a single antisymmetric form of
    axionic nature and a space-time with only two scale factors:  external,
    $a(t)$, and internal, $b(t)$. We assume the external 3-dimensional space
    to be isotropic (spherical, flat or hyperbolic) and seek globally
    regular solutions such that, at late times, $a(t)$ exhibits expansion
    while $b(t)$ and the dilaton $\varphi(t)$ tend to finite constant
    values. We will specify the requirement to the ``input'' parameters of
    the theory that lead to the existence of models with the desired
    properties. We shall see that they exist among closed (spherical) and
    flat models only for special values of the integration constants, in
    other words, require fine tuning. Unlike that, favourable hyperbolic
    models appear without fine tuning in a certain range of the input and
    integration constants and therefore seem to be much more realistic.

    There are many other arguments in favour of open cosmologies
    \cite{open_cos}. The observations are known to give the cosmological
    density factor $\Omega$ smaller or close to unity; meanwhile, the
    presently popular spatially flat cosmologies, most convenient for
    various calculations, require the precise equalty $\Omega=1$, actually a
    sort of fine tuning. It is much more probable that the real Universe at
    least slightly violates this special requirement.

\section{The model}

    Consider the action of $D$-dimensional gravity interacting with a
    dilatonic scalar field $\Phi$ and antisymmetric forms $F_s$, $F_p$,
    account the contributions from both the Neveu-Schwarz --- Neveu-Schwarz
    (NS-NS) and Ramond-Ramond (RR) sectors:
\beq  							\label{SJ}
    S_{\rm J} = \int d^D x \sqrt{g}\biggl\{
	\Phi \biggl[ \cR -\omega \frac{(\d\Phi)^2}{\Phi^2}
	-\sum_{s} \frac{\eta_s}{n_s!} F_s^2 \biggr]
	-\sum_{r} \frac{\eta_r}{n_r!} F_r^2\biggr\}
\eeq
    where $\cR$ is the scalar curvature, $g = |\det g_{MN}|$,
    $(\d\Phi)^2= g^{MN}\d_M\Phi \d_N \Phi$, $M,N = 0,\ldots, D-1$, $\omega$
    is a (Brans-Dicke type) coupling constant, $n_s$ and $n_r$ are the ranks
    of antisymmetric forms belonging, respectively, to the NS-NS
    and RR sectors of the effective action; for each $n$-form,
    $F_n^2 = F_{n,M_1\ldots M_n} F_n^{M_1\ldots M_n}$; the sign factors
    $\eta_s=\pm 1$ and $\eta_r =\pm 1$ depend on a particular field model.

    The action (\ref{SJ}) is written in the so-called Jordan conformal frame
    where the field $\Phi$ is nonminimally coupled to gravity. This form is
    actually obtained in the weak field limit of many underlying theories
    as the framework describing the motion of fundamental objects,
    therefore we will interpret the metric $g_{MN}$ appearing in (\ref{SJ})
    as the physical metric. Thus, if the fundamental objects are strings,
    one has in any dimension $\omega = - 1$, while in cases where
    such objects are $p$-branes, one finds \cite{duff}
\beq
    \omega = - \frac{(D-1)(p-1) -(p+1)^2}{(D-2)(p-1)- (p+1)^2},  \label{oDd}
\eeq
    where $p$ is the brane dimension and $D$ is the space-time dimension.
    The NS-NS sector of string theory predicts a Kalb-Ramond
    type field with $n_s = 3$; the type IIA superstring effective action
    contains RR terms with $n_r = 2,\ 4$, while type IIB predicts $n_r =
    3,\ 5$. The action (\ref{SJ}) may also represent the bosonic sectors of
    theories like 11-dimensional supergravity (where the dilaton is absent,
    and there is a 4-form gauge field), or 10-dimensional supergravity (there
    is a dilaton and a 3-form gauge field), or 12-dimensional ``field theory
    of F-theory'' \cite{khve}, admitting the bosonic sector of 11-dimensional
    supergravity as a truncation. The model \cite{khve} contains a dilaton
    and two $F$-forms of ranks 4 and 5; it admits electric 2- and 3-branes
    and magnetic 5- and 6-branes. The ``wrong" sign $\eta_r=-1$ is found in
    IIA* and IIB* supergravities, obtained with timelike T-duality from IIB
    and IIA theories, respectively, and also in the field limit of M*
    theory, appearing as a strong coupling limit of IIA* theory \cite{hukhu,
    hull}.

    The standard transformation
\beq
	g_{MN} = \Phi^{-2/(D-2)} \og_{MN}      		  \label{trans}
\eeq
    leads to a theory reformulated in the Einstein conformal frame, more
    convenient for solving the field equations:
\bear                                                     \label{SE}
    S_{\rm E} = \int d^D x \sqrt{g_{\rm E}}\biggl\{
	\ocR - \eta_{\omega} (\d\varphi)^2
	-\sum_{s} \frac{\eta_s}{n_s!} \e^{2\lambda_{s}\varphi}F_s^2
	-\sum_{r} \frac{\eta_r}{n_r!} \e^{2\lambda_{r}\varphi}F_r^2
	\biggr\}
\ear
    where all quantities are written in terms of the Einstein-frame metric
    $\og_{MN}$; $g_{\rm E} = |\det\og_{MN}|$;
    for the scalar field we have denoted
\beq
	\Phi = \e^{\varphi/\omega_1},\cm                    \label{Phi}
	\omega_1 = \sqrt{\biggl|\omega + \frac{D-1}{D-2}\biggr|};\cm
        \eta_\omega =\sign \biggl(\omega + \frac{D-1}{D-2}\biggr),
\eeq
    while the coupling constants $\lambda_s$ and $\lambda_r$ are
\bear
\label{lamd}
    \lambda_s \eql \frac{n_s -1}{\omega_1(D-2)}\cm \mbox{(NS-NS sector);}\nn
    \lambda_r \eql \frac{2n_r -D}{2\omega_1(D-2)}\cm \mbox{(RR sector).}
\ear

    The sign factor $\eom$ distinguishes ``normal'' theories ($\eom=+1$),
    such that the kinetic term of the $\varphi$ field in (\ref{SE}) has the
    normal sign corresponding to positive energy, from anomalous theories
    where this sign is ``wrong'' ($\eom = -1$). It should be noted that many
    theories with $D > 11$ involve $\eom = -1$.
    According to (\ref{Phi}),
\beq
    \frac{\eom}{\omega_1^2}
           = (D-2)\biggl[1 - \frac{(D-2)(p-1)}{(p+1)^2}\biggr]. \label{o1}
\eeq
    Evidently, under the condition $(D-2)(p-1) > (p+1)^2$
    we have $\eom =-1$. For $p=2,\ 5$ this happens when $D>11$, and for
    $p= 3,\ 4$ when $D > 10$.

    The following table gives the values of $\omega$ and $\eom/\omega_1^2$
    for some particular space-time and brane dimensions.

\def\z{\phantom{$-$}}
\begin{center}
\begin{tabular}{|c|c|l|l||c|c|l|l|}
\hline
$\wide \  D \ $ & \ $p$ \ & $\omega$ & $\eom/\oo^2$ &
		          \ $D$ \ & \ $p$ \ & $\omega$ & $\eom/\oo^2$ \\
\hline
 any    &  1  &    $-1$    &  $D-2$   & 12  &  2  &   $-2$   &   $-10/9$  \\
 10     &  2  &  \z 0      & \z 8/9   & 12  &  3  &  $-3/2$  &   $-5/2$   \\
 10     &  3  & \z$\infty$ & \z 0     & 12  &  4  &  $-8/5$  &   $-2$     \\
 10     &  4  &  \z 2      & \z 8/25  & 12  &  5  &   $-2$   &   $-10/9$  \\
 10     &  5  &  \z 0      & \z 8/9   & 12  &  6  &   $-6$   &   $-10/49$ \\
 10     &  6  &  $-4/9 $   & \z 72/49 & 12  &  7  & \z 1/2   & \z  5/8    \\
 11     &  2  & \z$\infty$ & \z 0     & 12  &  8  & $-4/11$  & \z 110/81  \\
 11     &  3  &    $-2$    & $-9/8$   & 14  &  2  & $-4/3 $  &$  -4     $ \\
 11     &  4  &    $-5/2$  & $-18/25$ & 14  &  6  & $-16/11$ &$ -132/49 $ \\
 11     &  5  & \z$\infty$ & \z 0     & 26  &  3  & $-17/16$ &$  -48    $ \\
 11     &  6  &  \z 1/4    & \z 36/49 & 26  &  4  & $-50/47$ &$-1128/25 $ \\
\hline
\end{tabular}
\end{center}

    Some comments are in order.
    First, the well-known result $\omega= - 1$ for strings ($p=1$) in any
    dimension is recovered. Second, one obtains $\omega=\infty$ for 2- and
    5-branes in 11 dimensions, which conforms to the absence of a
    dilaton in 11D supergravity that predicts such branes.
    Third, in 12 dimensions one has $\eta_\omega = -1$ for
    $p<7$, and such a theory \Ref{khve} does contain a pure
    imaginary dilaton: the $F$-forms of ranks 4 and 5 are coupled to a
    dilaton field $\varphi$ with the coupling constants $\lambda_1^2 =
    -1/10$ and $\lambda_2 = -\lambda_1$, respectively, while the product
    $\lambda\varphi$ is real. As is concluded in \Ref {khve},
    for $D>11$ ``imaginary couplings are exactly what is needed in order to
    make a consistent truncation to the fields of type IIB supergravity
    possible''. In our (equivalent) formulation, $\varphi$ and $\lambda$ are
    real and the unusual nature of the coupling is reflected in the sign
    factor $\eom$.

    Supersymmetric models with $D=14$ are also discussed \cite{bars,gavk},
    while $D=26$ is the well-known dimension for bosonic strings.

\section{Solutions}

    Let us specify the space-time structure and the Einstein-frame
    metric as
\bear
    \M \eql \R_u \times \M_0\times \M_1 \times \cdots \times \M_n,
	 \cm \dim \M_i = d_i,                                  \label{stru}
\\
    ds^2_{\rm E}                                               \label{dsE}
	\eql - \e^{2\alpha(u)} du^2 + \sum_{i=0}^{n} \e^{2\beta^i(u)}ds_i^2
\ear
    where $u$ is a time coordinate ranging in $\R_u\subset \R$ and
    $ds^2_i$ are the $u$-independent metrics of the factor spaces $\M_i$,
    assumed to be Ricci-flat for $i=1,\ldots,n$ whereas $ds_0^2$ in $\M_0$
    describes a space of constant curvature $K_0 = 0, \pm 1$, corresponding
    to the three types of isotropic spaces; $\M_0$ is thus interpreted as
    an external (observed) factor space.

    There is a diversity of exact solutions for the action (\ref{SE})
    without $L_m$ in space-times like (\ref{dsE}), discussed, in
    particular, in Refs.\,\cite{k1,k2,ivmel} (see also references therein).
    We will be only interested here in cosmological solutions for a very
    simple special case: a single antisymmetric form $F_{[d_0]}$ from the
    NS-NS or RR sector, having a single (up to permutations) nontrivial
    component $F_{1...d_0}$ where the indices refer to $\M_0$, and a single
    internal space $\M_1$, so that in (\ref{dsE}) $i=0,1$, and
    $\varphi = \varphi(u)$. Then the field equations are easily integrated.

    Let $u$ be a harmonic time coordinate for the metric (\ref{dsE}), so
    that $\alpha = d_0 \beta^0 + d_1\beta^1$.

    The $F$-form is magnetic-type; the Maxwell-like equations due to
    (\ref{SE}) are satisfied trivially while the Bianchi identity
    $dF=0$ implies
\beq
	F_{1...d_0} = Q \sqrt{g_0}, \cm Q = \const,           \label{Q}
\eeq
    where $g_0$ is the metric determinant corresponding to $ds^2_0$
    and $Q$ is a charge, to be called the {\it axionic charge\/} since the
    only nonzero component of $F$ can be represented in terms of a
    pseudoscalar axion field in $d_0+1$ dimensions. The remaining unknowns
    are $\beta^0$, $\beta^1$ and $\varphi$.

    In the Einstein equations $\ocR\mN - \half \delta\mN \ocR = T\mN$,
    written for the Einstein-frame metric (\ref{dsE}),
    the stress-energy tensor $T_M^N$ has the form
\beq                                                         \label{EMT'}
    \e^{2\alpha} T_M^N = -\half \etaF Q^2
    		\e^{2d_1\beta^1 +2\lambda\varphi}
			\diag (+1,\ [{-}1]_{d_0},\ [{+}1]_{d_1})
	-\half \eom \dot\varphi{}^2 \diag(+1,\ [{-}1]_{d_0+d_1})
\eeq
    where the first place on the diagonal belongs to $u$ and the
    symbol $[f]_d$ means $f$ repeated $d$ times; $\etaF = \pm 1$ is the
    sign factor of our $F$-form, originating from $\eta_s$ or $\eta_r$ in
    (\ref{SJ}) or (\ref{SE}).

    Due to the EMT property $T^u_u + T_z^z = 0$ (where $z$ belongs to
    $\M_0$), the corresponding Einstein equation has the Liouville form
    $\ddot\alpha - \ddot \beta^0 + K_0 (d_0-1)^2 \e^{2\alpha-2\beta^0}$,
    whence
\bearr
    \frac{1}{d_0-1}\e^{\beta_0-\alpha}                          \label{S}
    = S(-K_0,k,u) \eqdef \vars{\e^{ku},   & K_0 =0,\quad k\in\R;\\
			  k^{-1}\cosh ku, & K_0=1,\quad k>0;\\
			  k^{-1}\sinh ku, & K_0 =-1, \quad k> 0;\\
			  u,              & K_0 = -1,\quad k=0;\\
			  k^{-1}\sin ku,  & K_0 =-1, \quad k < 0, }
\ear
    where $k$ is an integration constant and one more constant is suppressed
    by a proper choice of the origin of $u$.
    \eq(\ref{S}) can be used to express $\beta^0$ in terms of $\beta^1$.

    It is helpful to consider the remaining unknowns
    as a vector $x^A = (\beta^1,\ \varphi)$
    in the 2-dimensional target space $\V$ with the metric
\beq
    (G_{AB}) = \pmatrix{        d\,d_1 & 0    \cr    	      \label{GAB}
	        	 	     0 & \eom \cr}, \cm
    (G^{AB}) = \pmatrix{    1/(d\,d_1) & 0    \cr
	        	 	     0 & \eom \cr},
		\cm   d \eqdef \frac{D-2}{d_0-1}.
\eeq
    The equations of motion then take the form
\bearr                                                       \label{eq-x}
    \ddot x{}^A = - \etaF Q^2 Y^A \e^{2y}
\\  \lal                                                         \label{int}
    G_{AB}\dot x^A \dot x^B + \etaF Q^2 \e^{2y} = \frac{d_0}{d_0-1}K,
    \cm
	 K = \vars {k^2\sign k, & K_0 = -1,\\
		    k^2,        & K_0 = 0, +1.}
\ear
    with the function $y(u) = d_1 \beta^1 + \lambda\varphi$, representable
    as a scalar product of $x^A$ and the constant vector $\vY$ in $\V$:
\beq
    y(u) = Y_A x^A, \cm  Y_A = (d_1,\ \lambda),  \cm
    		         Y^A = (1/d,\ \eom\lambda).       \label{YA}
\eeq
    \eq (\ref{int}) is a first integral of (\ref{eq-x}) that
    follows from the ${u \choose u}$ component of the Einstein equations.

    The simplest solution corresponds to $Q=0$ (scalar vacuum):
\beq
    \beta^1 = c^1 u + \uc^1, \cm                            \label{vac}
    \varphi = \cf u + \uc_\varphi,
\eeq
    where $c^1,\ \uc^1,\ \cf$ and $\uc_\varphi$ are integration constants.
    Due to (\ref{int}), the constants $c^A = (c^1,\ \cf)$ are related by
\beq                                                     \label{int-vac}
    c_A c^A = dd_1 (c^1)^2 + \eom \cf^2 = \frac{d_0}{d_0-1}K.
\eeq

    If $Q\neq 0$, \eqs (\ref{eq-x}) combine to yield an easily solvable
    (Liouville) equation for $y(u)$:
\beq
    \ddot y + \etaF Q^2 Y^2 \e^{2y}=0,                       \label{eq-y}
	\cm    Y^2 = Y_A Y^A = d_1/d + \eom \lambda^2.
\eeq
    This is a special integrable case of the equations considered, e.g., in
    Refs.\,\cite{k1, k2, ivmel}.

    Among the diverse solutions to (\ref{eq-y}) existing for different values
    of $Y^2$, we will choose, for our purposes, the solution for $Y^2 > 0$.
    One of the reasons is that even for $\eom=-1$ one has $Y^2 >0$ for
    fields from the NS-NS sector in any dimension and for fields from the RR
    sector if $D < 17$.  For $Y^2 > 0$, \eq (\ref{eq-y}) gives
\beq
    \e^{-y(u)} = \frac{|Q|Y}{h}\cosh [h(u+u_1)]            \label{y}
\eeq
    where $Y = |Y^2|^{1/2}$, $h >0 $ and $u_1$ are integration constants. The
    unknowns $x^A$ are expressed in terms of $y$ as follows:
\beq
    x^A = \frac{Y^A}{Y^2} y(u) + c^A u + \uc^A               \label{xA}
\eeq
    where the constants $c^A=(c^1,\ c_\varphi$ and $\uc^A =(\uc^1,\
    \uc_\varphi)$ satisfy the orthogonality relations
\beq
    c^A Y_A =0,\cm      \uc^A Y_A =0.                        \label{c-ort}
\eeq
    Finally, the constraint (\ref{int}) leads to one more relation
    among the constants:
\beq
    \frac{h^2}{Y^2} + c_A c^A = \frac{d_0}{d_0-1}K.         \label{int2}
\eeq

\section {Analysis of cosmological models}
\subsection {Prelimaries}

    In what follows, we put $d_0=3$, so that $d_1 = D-4$,
    and identify, term by term, the Jordan-frame metric \dsJ\ obtained in
    the above notations (\ref{trans}), (\ref{dsE}),
\beq
    \dsJ =                                                    \label{dsJ1}
     	  \exp\biggl[-\frac{2\varphi}{\omega_1(D-2)}\biggr]
    \biggl\{
 	  \frac{\e^{-d_1\beta^1}}{2S(-K_0,k,u)}
          \biggl[ \frac{-du^2}{4S^2(-K_0,k,u)} + ds_0^2\biggr]
 				+ \e^{2\beta^1} ds_1^2
 	          				\biggr\},
\eeq
    where the function $S(.,.,.)$ is defined in (\ref{S}), with the familiar
    form of the metric
\beq
    \dsJ = -dt^2 + a^2(t) ds_0^2 + b^2(t) ds_1^2,             \label{dsJ2}
\eeq
    so that $a(t)$ and $b(t)$ are the external and internal scale factors
    and $t$ is the cosmic time.

    To select nonsingular models, let us use the Kretschmann scalar ${\cal K}
    = R_{MNPQ}R^{MNPQ}$, which is in our case a sum (with positive
    coefficients) of squares of all Riemann tensor components
    $R_{MN}{}^{PQ}$. Thus as long as $\cal K$ is finite, all algebraic
    curvature invariants of this metric are finite as well. For the metric
    (\ref{dsJ2}) with $d_0=3$ one has (the primes denote $d/dt$):
\beq
    {\cal K} = 4\biggl[ 3\biggl(\frac{a''}{a}\biggr)^2      \label{Krch}
		    +d_1 \biggl(\frac{b''}{b}\biggr)^2
		+ 3 d_1 \biggl(\frac{a'b'}{ab}\biggr)^2\biggr]
	    +2\biggl[
	          6\biggl(\frac{K_0+{a'}^2}{a^2}\biggr)^2
		+ d_1(d_1-1)\,\frac{{b'}^4}{b^4}\biggr].
\eeq

    By (\ref{Krch}), ${\cal K} \to\infty$
    and hence the space-time is singular when $a\to 0$, $a\to\infty$,
    $b\to 0$ or $b\to\infty$ at finite proper time $t$.
    Accordingly, our interest will be in the asymptotic behaviour of the
    solutions at both ends of the range $\R_u = (u_{\min},\ u_{\max})$ of
    the time coordinate $u$, defined as the range where both $a^2$ and
    $b^2$ in (\ref{dsJ2}) are regular and positive. (Note that, as
    $t\to \pm\infty$, a singularity does not occur when $b(t)\to 0$, or
    $a\to 0$ in case $K_0=0$.) At any $u\in \R_u$ all the relevant functions
    are manifestly finite and analytical.  The boundary values $u_{\max}$
    and $u_{\min}$ may be finite or infinite; a finite value of $u_{\max}$
    or $u_{\min}$ coincides with a zero of the function (\ref{S}).

    Among regular solutions, of utmost interest are those in which
    $a(t)$ grows while $b(t)$ tends to a finite constant value as
    $t\to\infty$. Any asymptotic may on equal grounds refer to the
    evolution beginning or end due to the time-reversal invariance of the
    field equations. We will for certainty speak of expansion or inflation,
    bearing in mind that the same asymptotic may mean contraction
    (deflation).

    Let us now enumerate the possible kinds of asymptotics.

\medskip\noi
    {\bf Type I:} $u\to \pm \infty$, where
\beq
    dt^2 \sim \e^{(A-2k)|u|}, \cm
     a^2 \sim \e^{A|u|}, \cm                         \label{type1}
     b^2 \sim \e^{B|u|},
\eeq
    with $k>0$ and certain constants $A$ and $B$ depending on the solution
    parameters. A favourable asymptotic of $a(t)$ takes place when $A\geq
    2k$:
\begin{description}\itemsep -1.5pt
\item[(i)]
	$A>2k$: \ \ $t\to\infty$, \ \
	            $a\sim t^{A/(A-2k)}$\ (power-law inflation);
\item[(ii)]
	$A=2k$: \ \ $t\sim |u|\to \infty$,\ \
	            $a\sim \e^{kt}$\ (exponential inflation).
\end{description}
    A reformulation for $k<0$ is evident.
    The internal scale factor $b(t)$ tends to a finite limit if $B=0$,
    i.e., under a special condition on the model parameters (fine tuning).

\medskip\noi
    {\bf Type Ia:} a modification of type I when $k=0$, so that at
    $u\to\infty$
\beq
     dt^2 \sim u^{-3}\e^{Au}du^2,\cm
     a^2 \sim u^{-1}\e^{Au}du^2,\cm  b^2 \sim \e^{Bu}       \label{type1a}
\eeq
    If $A > 0$, we have, as desired, $t\to\infty$ and $a\to\infty$; the
    expansion may be called ``slow inflation" since it is only slightly
    quicker than linear: the derivative $da/dt \sim u$, which behaves
    somewhat like $\ln t$. If $A \leq 0$, then $a\to 0$ at finite $t$
    (singularity). As for $b(t)$, one may repeat what was said in case I.

\medskip\noi
    {\bf Type II:} $u\to 0$, where the function (\ref{S}) tends to zero,
    so that $S(-K_0,k,u) \sim u$, while other quantities involved are
    finite. In this case
\beq
    dt^2 \sim 1/u^3, \cm a^2 \sim 1/u, \cm b^2 \to \fin.    \label{type2}
\eeq
    According to (\ref{type2}), $t\to \pm \infty$, $a(t) \sim |t|$ (linear
    expansion or contraction), whereas both $b(t)$ and $\varphi(t)$ tend
    to finite limits since they do not depend on $S(-K_0, k, u)$.

    The dilaton $\varphi$ in all cases behaves like $\ln b(t)$, but, in
    general, with another constant $B$ in each particular solution.

    This exhausts the possible kinds of asymptotics for $Y^2 > 0$. Solutions
    with $Y^2 \leq 0$, which can emerge when $\eom = -1$ and/or with
    $\etaF=-1$, may have other asymptotics, but they are of lesser interest.

\subsection{Scalar-vacuum cosmologies}

    The scalar-vacuum models (\ref{dsJ1}), (\ref{vac}) depend on two input
    constants, $D$ (or $d_1=D-4$, or $d=(D-2)/2$) and $\omega$ (or
    $\omega_1$) and three integration constants $k,\ c^1,\ \cf$ related by
    (\ref{int-vac}); two more constants, $\uc^1$ and $\uc_{\varphi}$, only
    shift the scales in $\M_1$ and along the $\varphi$ axis and do not
    affect the qualitative behaviour of the models.

\medskip\noi
{\bf  Closed models, $K_0=+1$}. In this case in (\ref{dsJ1}) $S = k/\cosh
    ku,\ k>0$, hence the solution has two type I asymptotics at $u\to \pm
    \infty$, with $k>0$ and the following constants $A = A_{\pm}$:
\beq                                                            \label{Av+}
    A_{\pm} = -k
    		\mp \biggl[d_1 c^1 + \frac{\cf}{d\omega_1}\biggr],
\eeq
    so that at least at one of the asymptotics $A<0$ whence $a\to 0$ at
    finite $t$, a singularity. It is also easily seen that if $b(t) \neq
    \const$, it behaves as $\e^{Bu}$, $B=\const$, and tends to zero at one
    of the limits $u\to\pm\infty$.

\medskip\noi
{\bf Spatially flat models, $K_0=0$.} One has simply
\bear
     a^2(t) = \e^{Au},                                     \label{Av0}
     			\cm dt \sim \e^{(A-2k)u/2}du,
\ear
    where $A = - \cf/(d\omega_1) - d_1 c^1-k$,  $k\in\R$,
    and again $b^2(t) =\e^{Bu},\ B=\const$. Thus each of the scale factors
    is either constant, or evolves between zero and infinity, and $a=0$
    occurs at finite $t$.

\medskip\noi
{\bf Hyperbolic models, $K_0=-1$.} If $k>0$ [note that, when $\eom=1$, there
    is necessarily $k > 0$ due to (\ref{int-vac})], one has in (\ref{dsJ1})
    $S=k^{-1}\sinh ku$.  Hence the model evolves between a type I asymptotic
    at $u\to\infty$, with $A$ coinciding with $A_+$ in \eq (\ref{Av+}), and
    type II at $u=0$. Since type II is regular, a necessary condition for
    having a nonsingular model is $A\geq 2k$.

    To find out if and when it happens for $\eom = +1$, it is convenient
    to introduce, instead of the two constants $c^1$ and $\cf$ connected by
    (\ref{int-vac}), an ``angle'' $\theta$ such that
\beq
    -c^1 = \sqrt{\frac{3}{2dd_1}} k\,\cos\theta,\cm          \label{theta}
    -\cf = \sqrt{\frac{3}{2}} k\, \sin\theta.
\eeq
    The condition $A\geq 2k$ will be realized for a certain choice of the
    integration constants if $A_+$ given by (\ref{Av+}) has, as a function
    of $\theta$, a maximum no smaller than $2k$. An inspection shows that
    it happens if
\beq
    \omega_1^2 \leq 1/[d(6d - d_1)] = 1/[(D-1)(D-2)].        \label{omax}
\eeq
    This is the only example of a nonsingular (bouncing) vacuum model with
    $\eom=+1$.

    In case $k>0,\ \eom = -1$, a choice of $\cf$ and $c^1$ subject to
    (\ref{int-vac}) such that $A>2k$ is easily made for any $\omega_1$.

    For $\eom = -1,\ k=0$, the model evolves between type Ia and II
    asymptotics, where at the Ia end ($u\to\infty$)
\beq
	A = -d_1 c^1 - \cf/(d\omega_1),      \cm              \label{ABv-}
	B = 2 c^1 - \cf/(d\omega_1).
\eeq
    The necessary condition for regularity, $A > 0$, is satisfied for
    proper $c_1$ and $\cf$ which can be chosen without problems.

    In case $\eom = -1,\ k < 0$, the function $(\ref{S})$ is simply
    $|k|^{-1} \sin |k|u$, and the model has two type II asymptotics at
    adjacent zeros of $S$, say, $u=0$ and $u=\pi/|k|$. This model is
    automatically nonsingular for any further choice of integration
    constants.

    We conclude that among vacuum models only some hyperbolic ones can
    be nonsingular. For $\eom = +1$ in such a case $a(t)$ evolves from
    linear decrease to inflation, or from deflation to linear growth. Only
    in the latter case both $b(t)$ and $\varphi$ tend to finite limits as
    $t\to\infty$ without any fine tuning.

    For $\eom=-1$ there is a model interpolating between two
    asymptotics of the latter kind. Thus, as $t$ changes from $-\infty$ to
    $+\infty$, $a(t)$ bounces from linear decrease to linear increase
    (generically with a different slope) whereas $b(t)$ and $\varphi(t)$
    smoothly change from one finite value to another. The latter model
    exists for generic values of the integration constants.

\subsection{Cosmologies with an axionic charge}

    The solution contains, in addition to the input parameters $D$,
    $\omega$ and $\lambda$, three independent essential integration
    constants: the ``scale parameter'' $k$, the charge $Q$ and also $h$ and
    $\cf$ connected by (\ref{int2}); the constant $c^1$ is excluded by
    the first relation (\ref{c-ort})
\beq
    d_1 c^1 + \lambda \cf =0                                   \label{c1}
\eeq
    so that the quantity $c^A c_A$, appearing in (\ref{int2}), is expressed
    as $c^A c_A = \eom (d/d_1) \cf^2 Y^2$. The fourth constant, the
    ``shift parameter'' $u_1$, as well as $\uc^1$ and $\uc_\varphi$,
    connected by (\ref{c-ort}), are qualitatively inessential.

    Let us begin with ``normal'' models, $\eom = +1$.
    The solution (\ref{xA}) has the form
\bear
    \beta^1 (u) \eql \frac{1}{dY^2} y(u) + c^1 u +\uc^1,       \label{b1}
\\                                                             \label{f1}
    \varphi(u) \eql \frac{\lambda}{Y^2}y(u) + \cf u + \uc_\varphi.
\ear
    with $y(u)$ given by (\ref{y}). The form of (\ref{int2}) implies $k >
    0$ and suggests a notation similar to (\ref{theta}), namely,
\beq
    h = \sqrt{\frac{3}{2}} kY\cos\theta,
    \cm   						   \label{theta'}
    \cf = \biggl(1+\lambda^2\frac{d}{d_1}\biggr)^{-1/2}
    		\sqrt{\frac{3}{2}}k\sin\theta.
\eeq
    Let us now pass to the asymptotic description.

\medskip\noi
   {\bf Common asymptotic $u\to\infty$.} This asymptotic is common to all
   $K_0$ and belongs to type I with
\beq
    A = -k + h A_1 + \cf A_2, \cm
    B = h B_1 + \cf B_2                                     \label{AB1}
\eeq
    with the notations
\bear
    A_1 \eql \frac{1}{dY^2}\biggl(d_1 + \frac{\lambda\eom}{\oo}\biggr),
    	\cm
    A_2 = \lambda - \frac{1}{d\oo},                         \label{notAB}
    	\nn
    B_1 \eql \frac{1}{dY^2}\biggl(-2 + \frac{\lambda\eom}{\oo}\biggr),
    	\cm
    B_2 = -\biggl(\frac{2\lambda}{d_1} + \frac{1}{d\oo}\biggr).
\ear
    The condition $A\geq 2k$, or
\beq
      hA_1 + \cf A_2 \geq 3k,                                 \label{3k}
\eeq
    required for nonsingular models, may be realized for small $\omega_1$.
    Using the representation (\ref{theta'}), one can find, precisely as in
    the scalar-vacuum case, a condition under which $A_{\max}$, the maximum
    value of $A$ as a function of $\theta$, satisfies $A_{\max} \geq 2k$.
    Curiously, the coupling $\lambda$ drops away from the calculation, and
    the resulting condition coincides with (\ref{omax}).

    Meanwhile, $B$ can have any sign, therefore the asymptotic of $b(t)$
    is uncertain; however, by choosing the ratio $\cf/h$ (that is, by
    fine-tuning $\theta$) one can achieve $B=0$, so that $b(t)$ has a finite
    limit.

    The solution behaviour at the other end of the range $\R_u$ depends on
    $K_0$.

\medskip\noi
    {\bf Closed models.}
    For $K_0=+1,\ u\to\-\infty$ the asymptotic behaviour is
\beq
    a^2(t)\sim \e^{A'|u|},\cm
    b^2(t)\sim \e^{B'|u|},\cm dt\sim \e^{(A'-2k)|u|/2}du       \label{ab+}
\eeq
    where $A'$ and $B'$ coincide with $A$ and $B$ given by (\ref{AB1})
    with the replacement $\cf \to -\cf$ (or $\theta\to -\theta$).
    Therefore the description is the same up to the replacement $t\to -t$.
    When simultaneously $A\geq 2k$ and $A'\geq 2k$, one has a deflation
    $\to$ inflation transition for the scale factor $a(t)$, with generically
    different powers $A/(A-2k)$ and $A'/(A'-2k)$ at the contraction and
    expansion phases, and a regular bounce between them; at one end the
    evolution may be exponential. But, even if $b(t)\to \fin$ as
    $t\to\infty$, the evolution of $b(t)$ begins with $b=0$ or $b=\infty$
    unless the model parameters are further fine-tuned.

\medskip\noi
    {\bf Spatially flat models.}
    The case $K_0=0$ differs from $K_0=+1$ in that $k$ can have either
    sign, and when comparing the asymptotics $u\to\infty$ and
    $u\to -\infty$, one has to replace $k \to - k$. Hence, if we deal with a
    solution such that the Universe evolves in a power-law inflationary
    regime in the asymptotic $u \to \infty$, then, in the asymptotic $u \to
    - \infty$, we find a behaviour like $a(t) \sim |t|^{s},\ s<1$. So in the
    remote past there might be a subluminal contraction while in the remote
    future the model inflates. A necessary condition for this type of
    behaviour is again (\ref{3k}) (with $k$ replaced by $|k|$), leading to
    the requirement (\ref{omax}).

    Concerning $b(t)$, one only can repeat what was said about $K=+1$.

\medskip\noi
    {\bf Hyperbolic models.}
    The second asymptotic, $u\to 0$, is type II. Thus, provided $A > 2k$
    (see (\ref{AB1})), there occurs linear contraction of the Universe in
    the remote past and inflation in the remote future, or, in a
    time-reversed model, deflation in the remote past and linear expansion
    in the future. The latter opportunity seems especially attractive since
    both $\varphi$ (hence the string coupling) and $b(t)$ (hence the
    effective gravitational constant) tend to finite limits {\it
    automatically\/}, without any need for fine tuning.

    We see that even models where the kinetic terms of $\varphi$ and
    $F_{MNP}$ have both normal signs ($\eom = \etaF =1$), predict some
    nonsingular bouncing cosmologies.

    ``Anomalous'' models with $\eom =-1$, like their vacuum counterparts,
    bear some new features compared to $\eom =+1$, connected with the
    relation (\ref{int2}). Namely, \eqs (\ref{y}), (\ref{b1}), (\ref{f1})
    are still valid, but now we can have $c^A c_A \leq 0$. Therefore, first,
    the representation (\ref{theta'}) is no more valid, thus cancelling the
    restriction (\ref{omax}). Second, for $K_0=-1$, one can have now
    solutions with $k\leq 0$, with a different analytical form and different
    behaviour.

    As a result of the first of these circumstances, one obtains a
    bouncing behaviour of $a(t)$ in closed ($K_0=1$) and flat ($K_0=0$)
    models, having two type I asymptotics as described above, by simply
    choosing the integration constants $h$ and $\varphi_1$ from a proper
    range, without further restrictions on the input parameters (provided
    they lead to $\eom=-1$). However, fine tuning is still necessary for
    obtaining finite limits of $b(t)$ and $\varphi(t)$ at late times.

    The appearing hyperbolic models with $k=0$ interpolate between type II
    ($u\to 0$) and type Ia ($u\to\infty$) asymptotics (linear contraction
    $\to$ slow inflation or slow deflation $\to$ linear expansion), under
    the condition
\beq
	A = hA_1 +\cf A_2 >0,                                   \label{A0}
\eeq
    easily satisfied by choosing proper $h$ and $\cf$.

    There also appear bouncing models with $k < 0$ which qualitatively
    behave quite similarly to their vacuum counterparts, interpolating
    between two type II asymptotics.

\section{Concluding remarks}

    The bouncing mechanism discussed here works in a certain range of the
    free parameter of the model, the Brans-Dicke type constant $\omega$.
    Part of this favourable range, see \eq (\ref{omax}), corresponds to
    ``normal'' dilatonic fields with positive energy in the Einstein frame,
    and such bouncing models exist for all $D>4$.
    If we, however, ascribe the origin of $\omega$ to fundamental $p$-branes
    in the spirit of \Ref{duff}, it turns out that only ``anomalous''
    theories with $\eom=-1$ lead to bouncing models. [Indeed, in case $\eom
    =1$ \eq (\ref{o1}) gives $\omega_1^{-2} \leq D-2$, contrary to
    (\ref{omax}).] Moreover, such models only exist for $D > 11$, see
    the table in \sect 2 and the comments after it.

    The existence of a type II asymptotic of open models, making it possible
    to avoid fine tuning in getting a desired large $t$ asymptotic, is quite
    a general phenomenon for cosmologies with the structure (\ref{stru}). It
    actually follows from a ``stiff'' character of the stress-energy
    tensor (pressure in the external space is equal to energy density)
    that leads to \eq (\ref{S}) and is common to the dilaton and the axion
    in the Einstein frame.

    The present nonsingular vacuum and axionic models evidently cannot
    pretend to describe the full-time evolution of the Universe but rather
    the epoch of maximum contraction under the assumption that this stage
    is dominated by scalar-vacuum and axionic effects. Unlike other
    scenarios with a bounce at sub-Planckian scales, such as the brane gas
    \cite{brand99, brand00} and Pre-Big Bang \cite{gasvenez, gasperini}
    scenarios, no higher-order curvature terms are needed here to complement
    the model.  Furthermore, our models, with $b(t)$ and $\varphi(t)$
    tending to finite constant values, become effectively 4-dimensional at
    late times, with constant values of the effective gravitational constant
    and string tension, so the further evolution can be studied using
    conventional methods, in particular, an inflationary period can follow.
    Though, in our view, inflation is mostly needed for solving the problems
    of Big Bang cosmology emerging due to the existence of multiple causally
    disconnected regions, whereas bouncing cosmologies do not create such
    problems.

    In any bouncing model a separate problem is the origin of its
    initial state. We would note here that in some cases, as $t\to -\infty$,
    the dilaton field $\Phi = \e^{-\varphi/\oo}$ tends to zero, so that
    the string coupling parameter $g_s =1/\Phi$ diverges. This problem
    may probably be coped with due to the superstring dualities, mapping a
    strong coupling regime to a weak coupling regime. Such a difficulty is,
    however, absent in the above models interpolating between two type II
    asymptotics.

\Acknow
{We thank CAPES/COPLAG-ES and CNPq (Brazil) for partial financial support.}

\end{document}